\documentclass[conference,a4paper]{IEEEtran}
\usepackage{cite,graphicx,amsmath,amssymb,here}
\usepackage{multicol}
\usepackage{subfigure}
\usepackage{citesort}
\usepackage{fancyhdr}
\usepackage{dsfont}

\newtheorem{theorem}{Theorem}

\newtheorem{lemma}{Lemma}

\newtheorem{corollary}{Corollary}

\newtheorem{proposition}{Proposition}

\newtheorem{conjecture}{Conjecture}

\newtheorem{sketch}{Sketch of Proof}

\newcommand{\dis}{\stackrel{d}{\sim}}

\newcommand{\defeq}{\stackrel{\Delta}{=}}

\newcommand{\papertitle}{Spatial multiplexing with MMSE receivers: Single-stream optimality in ad hoc networks}

\begin{document}
\IEEEaftertitletext{\vspace{-400pt}}
\title{\papertitle}
\author{Raymond H. Y. Louie${}^{\natural}$,  Matthew R. McKay${}^{\dagger}$, Nihar Jindal${}^{\ddag}$, Iain B. Collings${}^{\S}$
\\
$ \natural${\small School of Electrical and Information Engineering, University of Sydney, Australia} \\
$\dagger${\small Department of Electronic and Computer Engineering, Hong Kong University of Science and Technology, Hong Kong} \\
${}^\ddag${\small Department of Electrical and Computer Engineering, University of Minnesota, Twin Cities, America} \\
${}^\S${\small ICT Centre, CSIRO, Sydney, Australia }
}
 \maketitle

\begin{abstract}
The performance of spatial multiplexing systems with linear minimum-mean-squared-error receivers is investigated in ad hoc networks. It is shown that single-stream transmission is preferable over multi-stream transmission, due to the weaker interference powers from the strongest interferers remaining after interference-cancelation. This result is obtained by new exact closed-form expressions we derive for the outage probability and transmission capacity.
\end{abstract}

\section{Introduction}

Multiple antennas can offer significant performance improvements in wireless
communication systems, by providing higher data rates and more reliable links. A practical
method which can achieve high data rates is to employ spatial multiplexing transmission in
conjunction with low complexity linear receivers, such as the minimum-mean-squared-error (MMSE) receiver. The MMSE receiver is particularly important as it uses its receive degrees of freedom (DOF) to optimally trade off strengthening the energy of the desired signal of interest and canceling unwanted interference, such that the signal-to-interference-and-noise ratio (SINR) is maximized.

In this paper, we investigate spatial multiplexing systems with MMSE receivers in ad hoc networks. The
transmitting nodes are spatially distributed according to a homogeneous Poisson point process (PPP) on a 2-D plane with density $\lambda$ (transmitting nodes per unit area), and send multiple data streams $N_t$ to their corresponding receiver. Besides corresponding to realistic
network scenarios, modeling the nodes according to a PPP has the benefit of allowing network
performance measures, such as the transmission capacity, to be obtained. The transmission capacity measures the maximum number of successful transmissions per unit area, assuming transmission at a fixed data rate, such that a target outage probability $\epsilon$ is attained.

To maintain a desired performance level for a fixed number of data streams per unit area $N_t \lambda$, a natural question arises whether it is preferable to have a high density of single-stream transmissions, or a low density of multi-stream transmissions. The main finding of this paper is that single-stream transmission is preferable when the optimal linear processing strategy, i.e.\ the MMSE receiver, is employed. This is due to the weaker interference powers from the strongest interferers remaining after interference-cancelation in single-stream transmission networks, compared to multi-stream transmission networks. This key result is facilitated by new exact closed-form expressions we derive for the outage probability and transmission capacity for arbitrary numbers of receive and transmit antennas.

Prior work on single-stream transmission with multiple receive antennas in ad hoc networks and Poisson distributed transmitting nodes include \cite{hunter08,huang08,govindasamy07,jindal09,jindal10,ali10}, where spectral efficiency and transmission capacity scaling laws were presented for different receiver structures.
In \cite{hunter08}, receive antennas are used for spatial diversity to increase the desired signal power, while in \cite{huang08}, receive antennas are used to cancel interference from the strongest interferer nodes. In \cite{govindasamy07}, MMSE receivers are used and the average spectral efficiency, a per-link performance measure, was obtained in the large antenna regime.  In \cite{jindal09,jindal10,ali10}, by using sub-optimal and MMSE linear receivers, the transmission capacity was shown to scale linearly with the number of receive antennas. In this paper, we extend these prior works to derive new outage probability and transmission capacity scaling laws for arbitrary number of data streams using MMSE receivers.

Multi-stream transmission with multiple receive antennas have been considered in \cite{Louie09_ICC,kountouris09,vaze09}. In \cite{Louie09_ICC,kountouris09}, spatial multiplexing systems were considered where receive antennas are used to cancel interference from the corresponding transmitter, but not the interferers. For these papers, the transmission capacity was shown to scale as\footnote{$f(x) = o(g(x))$ means $\lim_{x \to 0} \frac{ f(x)}{g(x)}=0$.} $o(\epsilon)$. A better scaling of $ o\left(\epsilon^{\frac{1}{L}}\right) $ was obtained in \cite{vaze09} by using sub-optimal receivers to cancel interference from the strongest $L-1$ interferers. This scaling result was used to show that single-stream transmission was preferable over multi-stream transmission when sub-optimal receivers are used \cite{vaze09}; in this paper we use a similar scaling result to show this is also true using optimal MMSE receivers.

\section{System Model}

We consider an ad hoc network comprising of transmitter-receiver
pairs, where each transmitter communicates to its corresponding
receiver in a point-to-point manner, treating
all other transmissions as interference. 
The transmitting nodes are distributed spatially according to a
homogeneous PPP of intensity $\lambda$ in $\mathds{R}^2$, and each receiving node is randomly placed at a distance $d_0$ away from its corresponding transmitter.

In this paper, we investigate network-wide performance. To
characterize this performance, it is sufficient to focus on a typical
transmitter-receiver pair, denoted by index 0, with the typical receiver located at the
origin. The transmitting nodes, with the exception of the
typical transmitter, constitute a marked PPP, which by Slivnyak's theorem, has the same distribution as the original PPP \cite{Stoyan95} (i.e., removing the typical transmitter from the transmit process has no effect). This
is denoted by $\Phi =\{ ( D_{\ell}, \mathbf{H}_{\ell}), \ell \in
\mathds{N} \}$, where $D_{\ell}$ and $\mathbf{H}_{\ell}$ model the location and channel matrix
respectively of the $\ell$th transmitting node with respect to
(w.r.t.) the typical receiver. The transmitted signals are attenuated
by a factor $1/r^{\alpha}$ with distance $r$ where $\alpha>2$ is the
path loss exponent.

We consider a spatial multiplexing system where each transmitting node sends $N_t$ independent data streams through $N_t$ different antennas to its corresponding receiver, which is equipped with $N_r$ antennas. Focusing on the $k$th stream, the received $N_r \times 1$ signal vector at the typical receiver can be written as
\begin{align}\label{eq:y}
\mathbf{y}_{0,k} &= \overbrace{\sqrt{\frac{1}{d_0^\alpha}} \mathbf{h}_{0,k} x_{0,k}}^{(a)} + \overbrace{\sqrt{\frac{1}{d_0^\alpha}} \sum_{q=1, q \neq k}^{N_t}
\mathbf{h}_{0,q} x_{0,q}}^{(b)} \notag \\
& \hspace{1cm} + \underbrace{\sum_{D_{\ell} \in \Phi}\sqrt{\frac{1}{|D_{\ell}|^\alpha}} \sum_{q=1}^{N_t} \mathbf{h}_{\ell,q} x_{\ell,q}}_{(c)}  +\mathbf{n}_{0,k}
\end{align}
where $x_{\ell,q}$ is the symbol sent from the $q$th transmit antenna of the $\ell$th transmitting node satisfying ${\rm E}[|x_{\ell,q}|^2]=P$, $\mathbf{h}_{\ell,q}$ is the $q$th column of\footnote{The notation $X \dis Y$ means that $X$ \emph{is
distributed as} $Y$.} $\mathbf{H}_{\ell} \dis
\mathcal{CN}_{N_r,N_t} \left(\mathbf{0}_{N_r \times N_t}, \mathbf{I}_{N_r}
\right)$ and $\mathbf{n}_{0,k}\dis \mathcal{CN}_{N_r, 1} \left(\mathbf{0}_{N_r \times 1}, N_0 \mathbf{I}_{N_r} \right)$ is the complex additive white Gaussian noise vector. We see in (\ref{eq:y}) that the received vector includes: (a) the desired data to be decoded, (b) the self interference from the typical transmitter and (c) the interference from the other transmitting nodes.

To obtain an estimate for $x_{0,k}$, we consider the use of MMSE linear receivers. The data estimate is thus given by $\hat{x}_{0,k}= \mathbf{h}_{0,k}^\dagger \mathbf{R}_{0,k}^{-1} \mathbf{y}_{0,k}$, from which the SINR can be written as
\begin{align}\label{eq:sinr}
{\rm SINR}_{0,k} = \frac{\gamma}{d_0^\alpha} \mathbf{h}_{0,k}^\dagger \mathbf{R}_{0,k}^{-1} \mathbf{h}_{0,k}
\end{align}
where
\begin{align}
\mathbf{R}_{0,k} = \frac{\gamma}{d_0^\alpha} \sum_{q=1, q \neq k}^{N_t}
\mathbf{h}_{0,q}\mathbf{h}_{0,q}^\dagger  + \gamma \sum_{ D_{\ell} \in \Phi} |D_{\ell}|^{-\alpha} \mathbf{H}_{\ell} \mathbf{H}_{\ell}^\dagger +  \mathbf{I}_{N_r} \;
\end{align}
and $\gamma=\frac{P}{N_0}$ is the transmit signal-to-noise ratio. We assume that each receiving node has knowledge of the corresponding transmitter channel $\mathbf{H}_0$ and the interference (plus noise) covariance matrix $\mathbf{R}_{0,k}$. 
The practicalities of this assumption are discussed in \cite{jindal10}.


\section{Outage Probability}

We consider the per-stream outage probability, defined for the $k$th stream as the probability that the mutual information for the $k$th stream lies below the data rate threshold $R_k$. At the receiver, the MMSE filter
outputs are decoded independently. We assume the data rate thresholds for all streams are the same and equal to $R$. The outage probability for each stream can thus be written as
\begin{align}
F_Z(z,\lambda) &= {\rm Pr}\left( {\rm SINR}\le z\right)
 \end{align}
where $z=2^R-1$ is the SINR threshold.  Note that we have dropped the subscript $k$ and 0 from the ${\rm SINR}$ term as the per-stream outage probability is the same for each stream at each receiving node.

Before presenting the outage probability, we first introduce some notation and concepts from number theory. The integer partitions of positive integer $k$ is defined as the different ways of writing $k$ as a sum of positive integers \cite{andrews04}. For example, the integer partitions of 4 are given by: i) 4, ii) 3+1, iii) 2+2, iv) 2+1+1 and v) 1+1+1+1. We denote $h(i,j,k)$ as the $i$th summand of the $j$th integer partition of $k$,
$|h(\cdot,j,k)|$ as the number of summands in the $j$th integer partition of $k$ and $|h(\cdot,\cdot,k)|$ as the number of integer partitions of $k$. For example, when $k=4$, we have  $h(2,3,4)=2$, $h(2,4,4)=1$, $|h(\cdot,3,4)|=2$ and $|h(\cdot,\cdot,4)|=5$.

We introduce non-repeatable integer partitions, which we define as integer partitions without any repeated summands. For example, the non-repeatable partitions of 4 are given by i) 4, ii) 3+1, iii) 2, iv) 2+1 and v) 1. We denote $g(i,j,k)$ as the number of times the $i$th summand of the $j$th non-repeatable integer partition of $k$ is repeated in the $j$th integer partition of $k$ and $|g(\cdot,j,k)|$ as the number of summands in the $j$th non-repeatable partition of $k$. For example, when $k=4$, we have $g(1,3,4)=2$, $g(1,5,4)=4$ and $|g(\cdot,3,4)|=1$.  Using these notations, we present a theorem for the outage probability\footnote{We note that the outage probability for the specific case where $N_t=1$ was recently independently derived in \cite{ali10}.}.

\begin{theorem}\label{the:outage_main}
The per-stream outage probability of spatial multiplexing systems with MMSE receivers is given by
\begin{align}\label{eq:outage_main}
& F_Z(z,\lambda) = 1 - \frac{e^{-\frac{z d_0^\alpha}{\gamma}} e^{-\Theta_{N_t} \lambda }}{(1+z)^{N_t-1}}    \sum_{p=0}^{N_r-1} \left(\sum_{\upsilon=1}^{N_r-p} \frac{\left(\frac{z d_0^\alpha}{\gamma}\right)^{\upsilon-1}}{(\upsilon-1)!} \right) \times \notag \\
&  \sum_{q=0}^{\min(p,N_t-1)}  \binom{N_t-1}{q} z^{q}    \sum_{j=1}^{|h(\cdot,\cdot,p-q)|} \Xi_{j,p-q} (-\Theta_{N_t}\lambda)^{|h(\cdot,j,p-q)|}
\end{align}
where
\begin{align}\label{eq:xi}
\Xi_{j,w} &= \frac{ \prod_{i=1}^{|h(\cdot,j,w)|}  \prod_{k=1}^{h(i,j,w)} \frac{\left(N_t-k+1\right)\left(k-1-\frac{2}{\alpha}\right)}{k\left(N_t+\frac{2}{\alpha}-k\right)}}{\prod_{\upsilon=1}^{|g(\cdot,j,w)|} g(\upsilon,j,w)!}
\end{align}
and
\begin{align}
\Theta_{N_t}= \frac{\pi \left(d_0^\alpha z \right)^{\frac{2}{\alpha}} \Gamma\left(N_t+\frac{2}{\alpha}\right) \Gamma\left(1-\frac{2}{\alpha}\right)}{\Gamma\left(N_t\right)} \; .
\end{align}
\end{theorem}
\begin{proof}
See the appendix.
\end{proof}

\begin{figure}[htbp]
\centerline{\includegraphics[width=1\columnwidth]{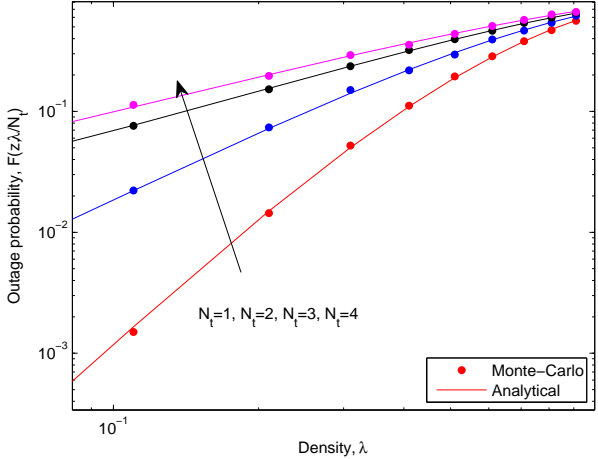}}
\caption{Outage probability vs.\ density for spatial multiplexing systems with MMSE receivers, and with $N_r=4$, $\alpha=4.6$, $z=0$ dB, $\gamma=20$ dB and $d_0=1$.} \label{fig:outage}
\end{figure}

For a fixed number of data streams per unit area, we can determine the optimal number of data streams used for transmission by considering the outage probability $ F_Z\left(z,\frac{\lambda}{N_t}\right)$. Fig.\ \ref{fig:outage} plots this outage probability vs.\ density for different number of antennas. The `Analytical' curves are based on (\ref{eq:outage_main}), and clearly match the `Monte-Carlo' simulated curves. We see that single-stream transmission always performs better then multi-stream transmission. In the next section, we analytically prove this is true using the transmission capacity framework for low outage probability operating values.

\section{Transmission Capacity}

We consider the transmission capacity, a measure of the number of successful transmissions per unit area, defined as
\begin{align}\label{eq:tc_def}
c(\epsilon) \defeq  N_t \lambda(\epsilon)(1 - \epsilon) R
\end{align}
where $\epsilon$ is the desired outage probability operating value and $\lambda(\epsilon)$ is the \emph{contention density}, defined as the inverse of $\epsilon=F_Z(z,\lambda)$ taken w.r.t.\ $\lambda$.  
The transmission capacity is given in the following lemma.
\begin{corollary}\label{corr_transmission_cap}
In the high SNR regime, the transmission capacity of spatial multiplexing systems with
MMSE receivers, subject to a low outage probability operating value,
is given by
\begin{align}\label{eq:transmission_cap_main}
c(\epsilon) = \frac{N_t R}{\Theta_{N_t} \Omega^{\ell} }  \epsilon^{\frac{1}{\ell}} + o\left( \epsilon^{\frac{1}{\ell}}\right)
\end{align}
where
\begin{align}
\ell = \left\lfloor \frac{N_r}{N_t} \right\rfloor \; ,
\end{align}
\begin{align}
\Omega 
&= \frac{1}{\ell!} -\frac{(-1)^\ell \sum_{q=0}^{N_t-1} \binom{N_t-1}{q} z^{q}  \sum_{p=\ell}^{N_r-1-q} \sum_{j \in \Psi_{p,\ell}} \Xi_{j,p}}{(1+z)^{N_t-1}}\; ,
\end{align}
$\Psi_{p,\ell}$ is the set of all integer partitions of $p$ with $\ell$ summands and $\lfloor \cdot \rfloor$ denotes the floor function.
\end{corollary}
\begin{proof}
The result is proven by taking a first order expansion of the outage probability in (\ref{eq:outage_main}) at high SNR around $\lambda=0$, followed by substituting the resultant expression into (\ref{eq:tc_def}). The full proof is omitted due to space limitations.
\end{proof}

By observing that the exponent of $\epsilon$ in (\ref{eq:transmission_cap_main}) is a decreasing function of the number of data streams, we see that for low outage probability operating values, the transmission capacity is maximized when only one data stream is used for transmission. This can be explained by considering the interference-cancelation properties of the MMSE receiver. As the MMSE receiver is the optimal linear processing strategy, the receive DOF is used to optimally trade off canceling the interference from the strongest interferers and strengthening the desired signals from the corresponding transmitter, such that the received SINR is maximized. 
The MMSE receiver is capable of completely canceling interference from both the corresponding transmitter and the strongest $k$ interferers if and only if $N_r > N_t -1 + k N_t$ \cite{gao98}, or equivalently $N_t < \frac{N_r+1}{k+1}$. The receiver can thus cancel interference from the $k-1$ strongest interferers if
\begin{align}\label{eq:condition_nearest}
\frac{N_r}{k+1} + \frac{1}{k+1}  \le N_t < \frac{N_r}{k} + \frac{1}{k} \; .
\end{align}
It can be shown that the value of $k$ satisfying the condition in (\ref{eq:condition_nearest}) corresponds to $k=\ell = \left\lfloor \frac{N_r}{N_t} \right\rfloor$. The MMSE receiver is thus capable of canceling interference from the $\ell-1$ strongest interferers.

As the transmission capacity increases with the number of strongest interferers whose interference is canceled, this implies the MMSE receiver will utilize the maximum possible DOF to cancel interference from the strongest interferers.  By noting that the receiver will require a minimum $N_t$ DOF to ensure the desired signals are received interference-free, the maximum $N_r-N_t$ DOF will be used to cancel interference from the strongest interferers. For single-stream transmission, the maximum (over all possible $N_t$) $N_r-1$ DOF are used to cancel interference. Thus single-stream transmission is preferable over multi-stream transmission as there are more strongest interferers whose interference are canceled. This implies that the interference powers originating from the strongest active interferers (whose interference is not canceled) are weaker for single-stream transmission than multi-stream transmission.

\begin{figure}[htbp]
\centerline{\includegraphics[width=1\columnwidth]{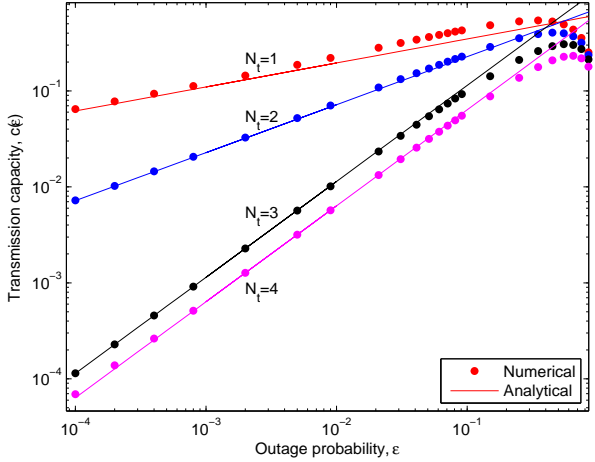}}
\caption{Transmission capacity vs.\ outage probability for spatial multiplexing systems with MMSE receivers, and with $N_r=4$, $\alpha=4.5$, $z=10$ dB, and $d_0=1$.} \label{fig:tc_outage}
\end{figure}

\begin{figure}[htbp]
\centerline{\includegraphics[width=1\columnwidth]{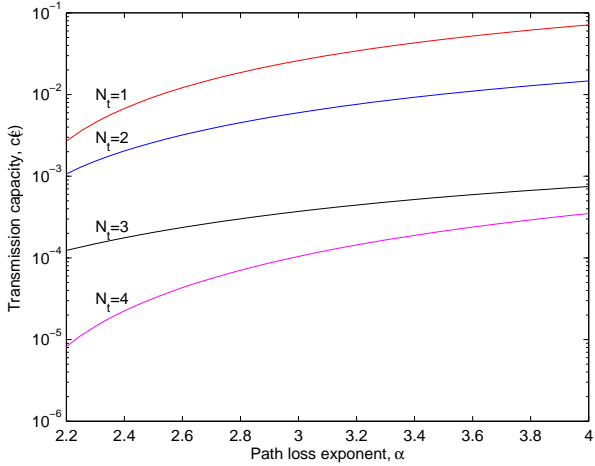}}
\caption{Transmission capacity vs.\ path loss exponent for spatial multiplexing systems with MMSE receivers, and with $N_r=4$, $z=15$ dB, $d_0=1$ and $\epsilon=0.001$.} \label{fig:tc_pathloss}
\end{figure}

Figs.\ \ref{fig:tc_outage} and \ref{fig:tc_pathloss} plot the transmission capacity vs.\ outage probability and path loss exponent respectively. We observe in both figures that the transmission capacity is a decreasing function of the number of transmit antennas for all outage probabilities and path loss exponents. In Fig.\ \ref{fig:tc_outage}, the `Analytical' curves are plotted using (\ref{eq:transmission_cap_main}), and closely match the `Numerical' curves for outage probabilities as high as $\epsilon=0.1$, which are obtained by numerically taking the inverse of $F_Z(z,\lambda)$ w.r.t.\ $\lambda$, and substituting the resulting expression into (\ref{eq:tc_def}).  Fig.\ \ref{fig:tc_outage} indicates that the optimality of single-stream transmission is not just applicable to small outage probability operating values, but the whole range of outage probabilities considered, i.e. $0.0001 \le \epsilon \le 0.8$. Fig.\ \ref{fig:tc_pathloss} indicates that the transmission capacity is an increasing function of the path loss exponent. This implies that for increasing path loss exponents, the positive effects of
the reduction in interference outweigh the negative effects of the reduction in desired signal strength between transmitter-receiver pairs.

\section{Conclusion}

The main takeaway message is that it is preferable to have a high density of single-stream transmissions than a low density of multi-stream transmissions using the optimal MMSE receiver in ad hoc networks. This is because the interference powers originating from the strongest interferers remaining after interference-cancelation are weaker for single-stream transmission than multi-stream transmission. This key result was obtained by new closed-form outage probability and transmission capacity expressions we derived for arbitrary numbers of transmit and receive antennas.

\begin{appendix}


The outage probability conditioned on $x_i=|D_i|^\alpha < a$, where $x_i$ are independent and identically uniformly distributed with $i=1,\ldots L$, is given by \cite{gao98}
\begin{align}\label{eq:cdf3}
F_{Z|x_1,\ldots,x_L}(z,\lambda) &= 1 - e^{-\frac{z d_0^\alpha}{\gamma}} \sum_{p=0}^{N_r-1}
\left(\frac{\sum_{\upsilon=1}^{N_r-p} \left(\frac{z d_0^\alpha}{\gamma}\right)^{\upsilon-1}}{(\upsilon-1)!} \right) \notag \\
& \hspace{1cm} \times  z^{p} d_0^{\alpha p}  I_p(x_1,\ldots,x_L,\lambda)
\end{align}
where
\begin{align}\label{eq:ip_general}
I_p(x_1,\ldots,x_L,\lambda) &=\frac{C_{p}(x_1,\ldots,x_L,\lambda)}{(1 + z)^{N_t-1} \prod_{i=1}^L (1
+ d_0^\alpha x_i^{-1} z)^{N_t}}
\end{align}
and $C_{p}(x_1,\ldots,x_L,\lambda)$ is the coefficient of $z^{p}$ in $(1 + d_0^{-\alpha} z)^{N_t-1}  \prod_{i=1}^L (1 + x_i^{-1} z)^{N_t}$.

To proceed, we average out the number of nodes, which follows a Poisson distribution, in $\mathcal{I}_p(x_1,\ldots,x_L,\lambda)$. This is given by
\begin{align}\label{eq:integral_general}
& {\rm E}\left[I_p(x_1,\ldots,x_L,\lambda) \right] 
= \frac{e^{-\lambda \pi a^{\frac{2}{\alpha}}}}{(1 + z)^{N_t-1} }  \sum_{L=0}^\infty \frac{  \left(\frac{2 \lambda \pi}{\alpha}\right)^L}{L!} \\
& \times \int_0^a \ldots \int_0^a C_{p}(x_1,\ldots,x_L,\lambda)  \prod_{i=1}^L \frac{x_i^{N_t+\frac{2}{\alpha}-1}}{ (x_i
+ d_0^\alpha  z)^{N_t}}  {\rm d} x_1 \ldots {\rm d} x_L \notag  \; .
\end{align}


To solve the integral in (\ref{eq:integral_general}), we are required to obtain an expression for $ C_{p}(x_1,\ldots,x_L,\lambda)$. To this end, it is convenient to first use the binomial series expansion to express $(1+ d_0^{-\alpha} z)^{N_t-1} \prod_{i=1}^L (1 + x_i^{-1} z)^{N_t}$ as
\begin{align}\label{eq:binomCp}
& (1+ d_0^{-\alpha} z)^{N_t-1} \prod_{i=1}^L (1 + x_i^{-1} z)^{N_t} 
= \\
& \sum_{q=0}^{N_t-1}  \sum_{q_1=0}^{N_t} \ldots \sum_{q_L=0}^{N_t} \left(\prod_{i=1}^L \binom{N_t}{q_i} x_i^{-q_i} \right)\binom{N_t-1}{q} \frac{ z^{q+\sum_{i=1}^L q_i}}{d_0^{\alpha q} } \notag \; .
\end{align}
We observe that the coefficient of $z^p$ in (\ref{eq:binomCp}) can be written as a sum of $\min(p+1,N_t)$ symmetric polynomials in $x_1^{-1},\ldots,x_L^{-1}$, corresponding to each term in the outer summation $ \sum_{q=0}^{N_t-1} $. These symmetric polynomials can be written as a sum of monomial polynomials, where the number of monomial polynomials is equal to the number of integer partitions of $p-q$, denoted by $|h(\cdot,\cdot,p-q)|$. As such, we can write the integral in (\ref{eq:integral_general}) as
\begin{align}\label{eq:integral_general2}
& {\rm E}\left[I_p(x_1,\ldots,x_L,\lambda) \right]= \frac{e^{-\lambda \pi a^{\frac{2}{\alpha}}}}{(1 + z)^{N_t-1} } \sum_{q=0}^{\min(p,N_t-1)}  \binom{N_t-1}{q}  \notag \\
& \times \frac{1}{d_0^{\alpha q}}  \sum_{L=0}^\infty \frac{  \left(\frac{2 \lambda \pi}{\alpha}\right)^L}{L!} \sum_{j=1}^{|h(\cdot,\cdot,p-q)|} \int_0^a \ldots \int_0^a \mathcal{M}_{j,p-q}(x_1,\ldots,x_L)   \notag \\
& \hspace{0.2cm} \times \prod_{i=1}^L \frac{x_i^{N_t+\frac{2}{\alpha}-1}}{ (x_i+ d_0^\alpha  z)^{N_t}}  {\rm d} x_1 \ldots {\rm d} x_L
\end{align}
where $\mathcal{M}_{j,p-q}(x_1,\ldots,x_L) $ is a monomial symmetric polynomial corresponding to the $j$th integer partition of $p-q$. We see that since the integral in (\ref{eq:integral_general2}) corresponding to the $j$th integer partition is symmetric w.r.t.\ $x_1,\ldots,x_L$, it is sufficient to solve this integral using only one monomial in $\mathcal{M}_{j,p-q}(x_1,\ldots,x_L) $ and multiply the resulting expression by the number of monomials in $\mathcal{M}_{j,p-q}(x_1,\ldots,x_L) $. We see in (\ref{eq:binomCp}) that the number of $x_i$ terms in each monomial comprising $\mathcal{M}_{j,p-q}(x_1,\ldots,x_L) $ is equal to the number of summands in the $j$th integer partition of $p-q$, denoted by $|h(\cdot,j,p-q)|$. Without loss of generality,  we thus focus on evaluating the integral of the monomial in  $x_1,\ldots,x_{|h(\cdot,j,p-q)|}$. By observing (\ref{eq:binomCp}), we finally make note that the coefficient of each monomial term is given by $\prod_{i=1}^{|h(\cdot,j,p-q)|} \binom{N_t}{h(i,j,p-q)}$, and that the number of monomials in $\mathcal{M}_{j,p-q}(x_1,\ldots,x_L) $ is given by $\Lambda_L=\frac{L!}{(L-|h(\cdot,j,p)|)! \prod_{\ell=1}^{|g(\cdot,j,p)|} g(\ell,j,p)!} $. Combining these facts, we can express (\ref{eq:integral_general2}) as
\begin{align}\label{eq:integral_general3}
&{\rm E}\left[I_p(x_1,\ldots,x_L,\lambda) \right]
= \frac{e^{-\lambda \pi a^{\frac{2}{\alpha}}}}{(1 + z)^{N_t-1} } \sum_{q=0}^{\min(p,N_t-1)}  \binom{N_t-1}{q} \notag \\
& \hspace{2cm} \times \frac{1}{d_0^{\alpha q}} \sum_{L=0}^\infty \frac{  \left(\frac{2 \lambda \pi}{\alpha}\right)^L}{L!} \int_0^a \ldots \int_0^a \mathcal{T}_{p,q}(x_1,\ldots,x_L)  \notag \\
& \hspace{2cm} \times \prod_{i=1}^L \frac{x_i^{N_t+\frac{2}{\alpha}-1}}{ (x_i+ d_0^\alpha  z)^{N_t}}  {\rm d} x_1 \ldots {\rm d} x_L
\end{align}
where
\begin{align}\label{eq:dp_general}
& \mathcal{T}_{p,q}(x_1,\ldots,x_L) \\
&= \sum_{j=1}^{|h(\cdot,\cdot,p-q)|} \Lambda_L  \prod_{i=1}^{|h(\cdot,j,p-q)|} x_i^{-h(i,j,p-q)} \prod_{k=1}^{h(i,j,p-q)} \frac{N_t-k+1}{k} \notag  \; .
\end{align}
To solve the integral in (\ref{eq:integral_general3}), it is convenient to define the following function:
\begin{align}
\mathcal{J}_\varsigma &=\int_0^a \frac{x^{N_t+\frac{2}{\alpha}-\varsigma-1}}{ (x
+ d_0^\alpha z)^{N_t}} {\rm d} x \\
 &=a^{N_t+\frac{2}{\alpha}-\varsigma} (d_0^\alpha z)^{-N_t} \Gamma\left(N_t+\frac{2}{\alpha}-\varsigma\right) \notag \\
 & \hspace{0.5cm} \times {}_2 \tilde{F}_1 \left(N_t,N_t+\frac{2}{\alpha}-\varsigma; N_t+\frac{2}{\alpha}-\varsigma+1; - \frac{a}{d_0^\alpha z} \right) \notag
\end{align}
where ${}_2 \tilde{F}_1(\cdot; \cdot; \cdot)$ is the regularized generalized Gauss hypergeometric function \cite{abramowitz70}. Now substituting (\ref{eq:dp_general}) into (\ref{eq:integral_general3}), we obtain
\begin{align}\label{eq:Igeneral}
& {\rm E}\left[I_p(x_1,\ldots,x_L,\lambda) \right]  
=\frac{e^{-\lambda \pi a^{\frac{2}{\alpha}}}}{(1 + z)^{N_t-1} }  \sum_{q=0}^{\min(p,N_t-1)} \binom{N_t-1}{q} \frac{1}{d_0^{\alpha q}} \notag \\
& \times \sum_{j=1}^{|h(\cdot,\cdot,p-q)|} \frac{\prod_{i=1}^{|h(\cdot,j,p-q)|}\mathcal{J}_{h(i,j,p-q)} \prod_{k=1}^{h(i,j,p-q)} \frac{N_t-k+1}{k}   }{\prod_{\upsilon=1}^{|g(\cdot,j,p)|} g(\upsilon,j,p)! \mathcal{J}_0^{|h(\cdot,j,p-q)|} } \Delta_L
\end{align}
where
\begin{align}
\Delta_L &= \sum_{L=|h(\cdot,j,p-q)|}^\infty \frac{  \left(\frac{2 \lambda \pi \mathcal{J}_0}{\alpha}\right)^L }{(L-|h(\cdot,j,p-q)|)!} \notag \\
&= \left(\frac{2 \lambda \pi \mathcal{J}_0}{\alpha}\right)^{|h(\cdot,j,p-q)|} e^{\frac{2 \lambda \pi \mathcal{J}_0}{\alpha}} \; .
\end{align}
To proceed, we take the limit as $a \to \infty$ in (\ref{eq:Igeneral}), since we are considering an infinite plane. It is thus convenient to note the following two limit functions,
\begin{align}\label{eq:limitfunc1}
& \lim_{a \to \infty} \exp\left(\frac{2 \lambda \pi}{\alpha}\mathcal{J}_0\right) e^{-\lambda \pi a^{\frac{2}{\alpha}}} \notag \\ & \hspace{1cm} = \exp\left(-  \frac{ \lambda \pi \left(d_0^\alpha z \right)^{\frac{2}{\alpha}} \Gamma\left(N_t+\frac{2}{\alpha}\right) \Gamma\left(1-\frac{2}{\alpha}\right)}{\Gamma\left(N_t\right)}  \right)
\end{align}
and
\begin{align}\label{eq:limitfunc2}
\lim_{a \to \infty} \frac{2 \lambda \pi}{\alpha}\mathcal{J}_\varsigma  
 = -\left(d_0^\alpha z \right)^{-\varsigma} \Theta_{N_t}\lambda \prod_{k=1}^{\varsigma} \frac{k-1-\frac{2}{\alpha}}{N_t+\frac{2}{\alpha}-k} \; .
\end{align}
Substituting (\ref{eq:limitfunc1}) and (\ref{eq:limitfunc2}) into (\ref{eq:Igeneral}), and substituting the resultant expression into (\ref{eq:cdf3}), we obtain the desired result.

\end{appendix}

\IEEEpeerreviewmaketitle
\bibliographystyle{IEEEtran}

\end{document}